\documentstyle[amssymb,prb,aps,psfig,epsf]{revtex}
\begin{document}
\newcommand{\btilde}{\mbox{$\tilde{b}$}}
\newcommand{\ktilde}{\mbox{$\tilde{\kappa}$}}
\newcommand{\gcheck}{\mbox{$\check{g}$}}
\newcommand{\sgn}{\mbox{${\rm sgn}$}}
\newcommand{\fcheck}{\mbox{$\check{f}$}}
\newcommand{\ghat}{\mbox{$\hat{g}$}}
\newcommand{\fhat}{\mbox{$\hat{f}$}}
\newcommand{\tauu}{\mbox{$\hat{\tau}_1$}}
\newcommand{\tauc}{\mbox{$\hat{\tau}_0$}}
\newcommand{\taud}{\mbox{$\hat{\tau}_2$}}
\newcommand{\taut}{\mbox{$\hat{\tau}_3$}}
\newcommand{\sigmac}{\mbox{$\hat{\sigma}_0$}}
\newcommand{\sigmau}{\mbox{$\hat{\sigma}_1$}}
\newcommand{\sigmad}{\mbox{$\hat{\sigma}_2$}}
\newcommand{\sigmat}{\mbox{$\hat{\sigma}_3$}}
\newcommand{\xx}{\mbox{$\partial_{xx}^2$}}
\newcommand{\ot}{\mbox{$\otimes$}}

\title{Manifestation of triplet superconductivity in superconductor-ferromagnet  structures}
\author{F.S.Bergeret $^{1 }$, A.F. Volkov$^{1,2}$ and K.B.Efetov$^{1,3}$}
\address{$^{(1)}$Theoretische Physik III,Ruhr-Universit\"{a}t Bochum,\\
 D-44780 Bochum, Germany\\
$^{(2)}$Institute of Radioengineering and Electronics of the Russian Academy
 of Sciences,\\ 103907 Moscow, Russia \\
$^{(3)}$L.D. Landau Institute for Theoretical Physics,\\
 117940 Moscow, Russia }

\maketitle

\begin{abstract}
We study proximity effects in a multilayered
superconductor/ferromagnet (S/F) structure with arbitrary relative
directions of the magnetization ${\bf M}$. If the magnetizations
of different layers  are collinear the superconducting
condensate function induced in the F layers has only a singlet
component and a triplet one with a zero projection of the total
magnetic  moment of the Cooper pairs on the ${\bf M}$ direction. In this  case the
condensate penetrates   the F layers over a short length $\xi_J$
determined by the exchange energy $J$. If the magnetizations ${\bf
M}$ are not collinear the triplet component has, in addition to
the zero projection, the projections $\pm1$. The latter component
is even in the momentum, odd in the Matsubara frequency and
penetrates  the F layers over a long distance that increases with
decreasing temperature and does not depend on $J$ ( spin-orbit
interaction limits this length). If the thickness of the F layers
is much larger than $\xi_J$, the Josephson coupling between
neighboring S layers is provided only by the triplet component, so
that a new type  of superconductivity arises in the transverse
direction of the structure. The Josephson critical current is
positive (negative) for the case of a positive (negative)
chirality of the vector  ${\bf M}$. We demonstrate that this type
of the triplet condensate can be detected also by measuring the
density of states in F/S/F structures.

\end{abstract}

\section{Introduction}

Multilayered superconductor/ferromagnet ($S/F)$
structures are under an intensive study now
(for a recent review see e.g. \cite{Proshin}%
). The interest in such systems originates from a possibility to
find new physical phenomena as well from the hope to construct new
devices based on these structures. Although a ferromagnet F
attached to a superconductor S is expected to suppresses the order
parameter in S, under certain conditions superconductivity and
ferromagnetism may coexist and exhibit interesting phenomena.

 One of them is a nonmonotonic dependence of the critical temperature
$T_{c}$ of the superconducting transition in $S/F$ multilayered
structures on the thickness $d_{F}$ of the ferromagnetic layers.
Theory of this effect has been developed in Refs.\cite{TheorTc},
and experimental results  have been presented in
Refs.\cite{ExperTc}.

Another interesting phenomenon is a $\pi -$state that can be
realized in $SFS$ Josephson junctions. It was shown
\cite{PiContact} that for some values of parameters (such as the
temperature $ T $, the thickness $d_{F}$, the exchange energy $J$)
the lowest Josephson energy corresponds not to the zero
phase difference $\varphi $, but to $%
\varphi =\pi $ (negative Josephson critical current $I_{c}$).
Detailed theoretical studies of this effect have been presented in
many papers \cite{a,BVE3}. The $\pi$-state has been observed
experimentally in Refs.\cite{Ryazan}.

Later it was discovered that the critical current $I_{c}$ in
Josephson junctions with ferromagnetic layers is not  necessarily 
suppressed by the exchange interaction and it may even
be enhanced. Such an enhancement of $I_{c}$ has been demonstrated
by the present authors on a simple model of a $SF/I/FS$ junction,
where $I$ stands for a thin insulating layer \cite{BVE1}. It was
shown for thin S and F layers that at low temperatures the
critical current $I_{c}$ in a $SF/I/FS$ junction may become even
larger than in the absence of the exchange field (i.e. if the $F$
layers are replaced by $N$ layers, where $N$ is a nonmagnetic
metal). More detailed calculations of $I_{c}$ (for arbitrary $S/F$
interface transmittance) for this and similar junctions have been
performed later in Refs.\cite {Enhanc}.
% It was shown that the current $I_{c}$ may both increase or decrease
%with increasing exchange energy $J$.

{\ Properties of superconductors in S/F structures may change not only due to the proximity effect  but also due to the long-range magnetic interaction. A spontaneous creation of vortices caused by the magnetic interaction has been predicted in a S/I/F system ( I is an insulating layer)\cite{erdin}.} 
In most papers on $S/F$ structures the case of collinear (parallel
or antiparallel) orientations of the magnetization ${\bf M}$ was
considered. If the magnetization vector ${\bf M}$ is not constant
in space, as in a domain wall,  or if the orientations of ${\bf
M}$ in different $F$ layers are not collinear to each other, a
qualitatively new and interesting effect occurs. For example, if a
ferromagnetic wire is attached to a superconductor, a domain wall
in the vicinity of the interface can generate a triplet component
of the superconducting condensate \cite{BVE2} (a similar case was
analyzed in a later work \cite{Swed}).

 The existence of the triplet component ($TC$) has far reaching consequences. It is
well known that the singlet component ($SC$) penetrates into  a ferromagnet over the length $\xi _{J}=%
\sqrt{D_{F}/J}$, where $D_{F}$ is the diffusion coefficient in
$F$. In contrast, it was shown that even for $J>>T$ the $TC$
penetrated $F$ over a much longer distance $\xi
_{T}=\sqrt{D_{F}/2\pi T}.$ This long-range penetration of the $TC$
might lead to an increase of the conductance of the $F$ wire if
the temperature is lowered below $T_{c}$\cite{BVE2,Swed}.

In this paper we consider a multilayered $S/F$ structure. Each $F$
layer has a constant magnetization ${\bf M}$ but the direction of
the ${\bf M}$ vector varies from layer to layer. We show that, in
this case, the triplet component of the superconducting condensate
is also generated and it penetrates  the $F$ layers over the long
length $\xi _{T} $ that does not depend on the large exchange
energy $J$ at all.

If the thickness of the F layers $d_{F}$ is much larger than $\xi
_{J}$, then the Josephson coupling between adjacent $S$ layers and,
therefore, superconductivity in the transverse direction is due to
the $TC$. In the vicinity of the $S/F$ interface the amplitudes of
the $SC$ and $TC$ may be comparable but, unlike the $TC$, the $SC$
survives in $F$ only over the short distance $\xi _{J}$ from the
$S/F$ interface. In other words, in the multilayered $F/S$
structures with a non-collinear magnetization orientation, a new
type of superconductivity arises. The non-dissipative current
within the layers is due to the s-wave singlet superconductivity,
whereas the transversal supercurrent across the layers is due to
the s-wave, triplet superconductivity.

It is important to emphasize (see Ref.\cite {BVE2}) that the $TC$
in this case differs from the $TC$ realized in the superfluid
He$^{3}$ and, for example, in materials like Sr$_2$RuO$_4$ \cite
{SrRuO}. The triplet-type superconducting condensate we predict
here is symmetric in momentum and therefore is insensitive to
non-magnetic impurities. It is odd in frequency and is called
sometimes odd superconductivity.

This type of the pairing has been proposed by Berezinskii in 1975
\cite{Berez} as a possible candidate for the mechanism of
superfluidity in He$^{3}.$ However, it turned out that another
type of pairing was realized in He$^{3}$: triplet, odd in momentum
$p$ (sensitive to ordinary impurities) and even in the Matsubara
frequencies $\varpi $. Attempts to find conditions for the
existence of the odd superconductivity were undertaken later in
several papers in connection with the pairing mechanism in high
$T_{c}$ superconductors \cite{Balats} (in Ref. \cite{Balats} a
singlet pairing odd in frequency and  in the momentum was
considered). {\ It is also important to note that while the symmetry of the order parameter $\Delta$ in Refs.\cite{SrRuO,Berez,Balats} differs from that of the BCS order parameter, in our case $\Delta$ is nonzero only in the S layers and is of the BCS type. It is determined by the amplitude of the singlet component. Since the triplet and singlet components are connected which each other, the TC affects $\Delta$ in an indirect way.}

Therefore the type of superconductivity analyzed in our paper
complements the three known types of superconductivity: s-wave and
d-wave  singlet superconductivity that occur in ordinary
superconductors and  in high $T_{c}$ superconductors respectively,
and the p-wave superconductivity with triplet pairing  observed in
Sr$_2$RuO$_4$.

In addition,  the new type of the triplet  superconductivity
across the $S/F$ layers shows another interesting property related
to the chirality of the magnetization ${\bf M}$. If the angle of
the magnetization rotation $2\alpha$ across the  $S_{A}$ layer
(see Fig.\ref{geom2}) has the same sign as the angle of the ${\bf
M}$ rotation  across the $S_{B}$
layer, then the critical Josephson current $I_{c}$ between $S_{A}$ and $%
S_{B}$ is positive. If these angles have different signs, then the
critical current $I_{c}$ is negative and $\pi - $ state is
realized (in this case spontaneous supercurrents arise in the
structure). This negative Josephson coupling, which is caused by
the $TC$ and depends on chirality, differs from that analyzed in
Ref. \cite{PiContact}. Depending on the chirality an ''effective''
condensate density in the direction perpendicular to the layers
may be both positive and negative.
%%%%%%%%%%%%
{\  We note that a dependence of the Josephson current on chirality has also 
been obtained in Ref.\cite{Kulic2}. The authors of Ref.\cite{Kulic2} considered two 
magnetic superconductors $S_m$  with  spiral magnetization, separated by 
a thin insulating layer I. In the latter case the TC exists  in the bulk superconductors together with the SC ( and they cannot be separated) and the Josephson current depends on the chirality of the spiral structures. The main  
difference between our system and the system considered by the authors of Ref.\cite{Kulic2}  is that in our case only the long-range TC survives in the F 
layers whereas in the $S_mIS_m$  junction both the SC and TC exist simultaneously. Therefore in the case of a collinear alignment of {\bf M}, the Josephson 
coupling (and triplet superconductivity in the transverse direction) 
disappears in our system, whereas it remains in the  $S_mIS_m$  system.}
%%%%%%%%%%%%%%%

Another possible detection of the TC in the S/F structures may be
achieved by measuring the density of states (DoS) in a F/S/F
trilayer (see Fig. \ref{geom1}). We will see that the long-range
TC causes a measurable change of the local  DoS at the outer side
of the F layers even if $d_F$ is much larger than $\xi_J$.

The plan of this paper is as follows. In the next section we make
some preliminary remarks concerning the TC in S/F structures.  We
consider a three-layer $FSF$ structure and calculate the
condensate function in this structure. We show that the amplitude
of the $TC$ is proportional to $\sin \alpha $ and its long-range
part is an odd function of the Matsubara frequency $\varpi $ (the
$SC$ is an even function of $\varpi $), where $\pm \alpha $ is the
angle between the $z$-axis and the magnetization in the right
(left) $F$ layers. We discuss properties of the $TC$ and calculate
the
 DOS related to it. In Sec. 3 we calculate the
Josephson current between adjacent $S$ layers and discuss its
dependence on the chirality of the magnetization variation in the
system. In Sec. 4 we take into account spin-orbit interactions and
study the effect of this interaction on the $TC$. In the
conclusion we discuss the obtained results and possibilities of an
experimental observation of the predicted
effects. The odd triplet superconductivity in $%
F/S$ structures was first predicted by the present authors in a
short paper where the case of small angles $\alpha $ and of a
perfect $F/S$ interface was considered \cite{VBE}.

%%%%%%%%%%%%%%%%%%%%%%%%%%%%%%%

\section{The condensate function in a  F/S/F sandwich}

\begin{figure}
\epsfysize= 5cm \vspace{0.2cm}
\centerline{\epsfbox{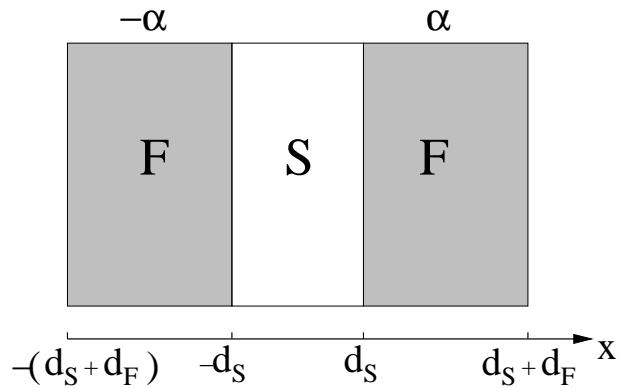}} \vspace{0.2cm}
\caption{The F/S/F trilayer. The magnetizations vectors in the F
layers make an angle $\pm\alpha$ with  the $z$-axis, respectively}
\label{geom1}
\end{figure}

In order to get a better understanding of the properties of the superconducting
condensate in the presence of the ferromagnetic layers, we consider
in this section a simple case of a trilayered F/S/F structure (see
Fig.\ref{geom1}). Generalization to a multilayered structures is
of no difficulties and will be done in the next section.

In the most general case, when the magnetization vectors ${\bf M}$
of the F-layers are non-collinear, the electron Green functions
are $4 \times 4$ matrices in the  particle-hole$\otimes
$spin-space\cite{foot}. The $4 \times 4$ matrix Green functions have been
introduced long ago \cite{VGL} and used in other papers\cite{Maki}. Later on they were used in Ref.\cite{Kulic} for a
description of magnetic superconductors with a rotating
magnetization.

A very convenient way for the study of  proximity effects is the
method of quasiclassical Green's functions \cite{eilen,lo,usadel}.
Equations for the quasiclassical Green's functions have been
generalized recently to the case of a non-homogeneous exchange
field (magnetization) ${\bf M}$ \cite{BEL}.

Following the notation of Ref.\cite{BVE3} we represent the
quasiclassical Green functions in the form
\begin{equation}
\mbox{$\check{g}$}=g_{ss^{\prime }}^{nn^{\prime}}=\frac{1}{\pi }\sum_{n^{\prime\prime}}(%
\mbox{$\hat{\tau}_3$})_{nn^{\prime \prime }}\int d\xi _{p}\left\langle \psi
_{n^{\prime \prime}s}(t)\psi _{n^{\prime }s^{\prime }}^{+}(t^{\prime })\right\rangle \;,
\label{gf}
\end{equation}
where the subscripts $n$ and $s$ stand for the elements in the
particle-hole and spin space, respectively, and $\mbox{$\hat{\tau}_3$}$ is
the Pauli matrix. The field operators $\psi _{ns}$ are defined as
$\psi _{1s}=\psi _{s}$ and $\psi _{2s}=\psi _{\bar{s}}^{+}$ (
$\bar{s}$ denotes the opposite to $s$ spin direction).

The elements of the matrix $\mbox{$\check{g}$}$  diagonal in the
particle-hole space (i.e proportional to $\mbox{$\hat{\tau}_0$}$ and $%
\mbox{$\hat{\tau}_3$}$) are related to the normal Green's function,
while the off-diagonal elements (proportional to $\mbox{$\hat{\tau}_1$}$ and $%
\mbox{$\hat{\tau}_2$}$) determine the superconducting condensate function $%
\mbox{$\check{f}$}$. In the case under consideration the matrix (\ref{gf})
can be expanded in the Pauli matrices in the particle-hole space ($%
\mbox{$\hat{\tau}_0$}$ is the unit matrix):
\begin{equation}
\mbox{$\check{g}$}=\mbox{$\hat{g}$}_{0}\mbox{$\hat{\tau}_0$}+\mbox{$\hat{g}$}%
_{3}\mbox{$\hat{\tau}_3$}+\mbox{$\check{f}$}\;,  \label{gfunc}
\end{equation}
where the condensate function is given by
\begin{equation}
\mbox{$\check{f}$}=\mbox{$\hat{f}$}_{1}i\mbox{$\hat{\tau}_1$}+%
\mbox{$\hat{f}$}_{2}i\mbox{$\hat{\tau}_2$}\;.  \label{ffunc}
\end{equation}
The functions $\mbox{$\hat{g}$}_{i}$ and $\mbox{$\hat{f}$}_{i}$
are matrices in the spin-space. In the case under consideration the
matrices $\mbox{$\hat{f}$}_{i}$ can be represented in the form
\begin{eqnarray}
\mbox{$\hat{f}$}_{2}(x) &=&f_{0}(x)\mbox{$\hat{\sigma}_0$}+f_{3}(x)%
\mbox{$\hat{\sigma}_3$} \\
\mbox{$\hat{f}$}_{1}(x) &=&f_{1}(x)\mbox{$\hat{\sigma}_1$}
\end{eqnarray}
This follows from the equation that determines the Green's function (see below).

Let us discuss briefly properties of the condensate matrix
function \mbox{$\check{f}$}. According to the definitions of the
Green's functions, Eq. (\ref{gf}), the functions $f_{i}(x)$ are
related to following correlation functions
\begin{eqnarray}
f_{3} &\sim &\left\langle \psi _{\uparrow }
\psi _{\downarrow }\right\rangle
-\left\langle \psi _{\downarrow }
\psi _{\uparrow }\right\rangle \;,
\nonumber \\
f_{0} &\sim &\left\langle \psi _{\uparrow }
\psi _{\downarrow }\right\rangle
+\left\langle \psi _{\downarrow }
\psi _{\uparrow }\right\rangle \;, \\
f_{1} &\sim &\left\langle \psi _{\uparrow }
\psi _{\uparrow }\right\rangle
\sim \left\langle \psi _{\downarrow }
\psi _{\downarrow }\right\rangle \;.
\nonumber
\end{eqnarray}
%%%%%%%%%%%%%%%
%%%%%%%%%%%%%%%%%
The function $f_{3}$ describes the $SC$, while the functions
$f_{0}$ and $f_{1}$ describe the $TC$ (see for example
Ref.\cite{legg}). The function $f_{0}$ is proportional to the zero
projection of the triplet magnetic moment of the Cooper  pairs on the z-axis,
whereas the function $f_{1}$ corresponds to the projections $\pm
1$.

It is important that in the absence of an exchange field ${\bf J}$
(or magnetization ${\bf M}$) acting on spins, the $SC$, i.e. the
function $f_{3}$, exists both in the superconducting and normal
(non-magnetic) layers. If $J$ is not equal to zero but is uniform
in space and directed along the z-axis, then the part $f_{0}$ of
the $TC$ arises in the structure.

However, both the functions $f_{3}$ and $f_{0}$ decay very fast in
the ferromagnet (over the length $\xi_{J}$). The singlet component
decays because a strong magnetization makes the spins of a pair be
parallel to each other, thus destroying the condensate. The
triplet component with the zero projection of the magnetic moment
is also destroyed because it is more energetically favorable for
the magnetic moment to be parallel to the magnetization.

On the other hand, the structure of the matrix
$\mbox{$\check{f}$}$ (the functions $\mbox{$\hat{f}$}_{i}$) depends
on the choice of the $z$-axis. If the uniform magnetization ${\bf
M}$ is directed not along the $z$-axis (but,say, along the
$x$-axis), terms like $\mbox{$\hat{f}$}_{1}i\mbox{$\hat{\tau}_1$}$
inevitably appear in the condensate function (see for example,
Ref.\cite{Kulic} where such a term was obtained even at $Q=0$, $Q$ is
the wave vector of a spiral magnetic structure). However, the
condensate component corresponding to this term penetrates  the $F$
layer over the short distance $\xi _{J}$ only.

Therefore, we can conclude that the presence  of  terms like
$\mbox{$\hat{f}$}_{1}i\mbox{$\hat{\tau}_1$}$ in the condensate
function does not necessarily mean that the $TC$ penetrates 
the $F$ layer over the long distance  $\xi _{T}$. Actually,
long-range effects arise only if the direction of the vector ${\bf
M}$ varies in space. If the magnetization has different directions
in neighboring $F$ layers, then not only $f_{0}$ but also $f_{1}$
arise in the system and both  functions penetrate  the
ferromagnetic layer over a long distance $\xi _{T}$.

In order to find the Green's function $\gcheck$, we consider the
diffusive case when the Usadel equation is applicable. This
equation can be used provided the condition $J\tau\ll1$ is
satisfied ( $\tau$ is the momentum relaxation  time). Of course,
this condition can hardly be satisfied for strong ferromagnets
like $Fe$ and  in this case one should use a more general
Eilenberger equation for a quantitative computation. However, the
Usadel equation may give qualitatively reasonable results even in
this case.

The Usadel equation is a nonlinear equation for the 4$\times$4
matrix Green's function $\gcheck$ and can be written as
\begin{equation}
D\partial _{x}\left(\gcheck\partial_x\check{g}\right)
-\omega \left[\taut\sigmac,\gcheck\right]
+iJ\left\{\left[\taut\sigmat,\gcheck\right]
\cos\alpha(x)+\left[\tauc\sigmad,\gcheck\right]
\sin\alpha(x)\right\}=-i\left[\check{\Delta},\gcheck\right]\; .
\label{Usad}
\end{equation}
In the S layer $D=D_S$, $J=0$, $\check{\Delta}=\Delta
i\taud\sigmat$ (the phase of $\Delta$ is chosen to be zero). In
the F layers $D=D_F$, $\alpha(x)=\pm\alpha$ for the right (left)
layer and $\Delta=0$. Eq. (\ref{Usad}) is complemented by the
boundary conditions at the S/F interface\cite{Zaitsev}
\begin{eqnarray}
\gamma\left(\gcheck\partial_x\check{g}\right)_F&=
& \left(\gcheck\partial_x\check{g}\right)_S,\;\;
x=\pm d_S \label{bc1} \\
2\gamma_b\xi_J\left(\gcheck\partial_x\check{g}\right)_F &=
&\pm \left[\gcheck_S,\gcheck_F\right],\;\; x=\pm d_S\; ,  \label{bc2}
\end{eqnarray}
where $\gamma=\sigma_F/\sigma_S$, and $\sigma_{S,F}$ are the
conductivities of the F and S layers, $\gamma_b=\sigma_FR_b/\xi_J$
is a coefficient characterizing the transmittance of the S/F
interface with resistance per unit area $R_b$.

If linearized, the Usadel equation can be solved analytically
rather easily. The linearization may be justified in the two
limiting cases: a) T is close to the critical temperature of the
structures $T_c^*$ (the latter can be different from the critical
temperature of the bulk superconductor $T_c$), and b) the
resistance of the S/F interface $R_b$ is not small. In the latter
case the condensate function in the $S$ layer is weakly disturbed
by the F film and the function $f_3$ in Eq. (\ref{ffunc}) can be
represented in the form
\begin{equation}
f_3(x)=f_S+\delta f_3(x),\;\;\; |x|<d_S\; ,
\label{f3}
\end{equation}
where $f_S=\Delta/iE_\omega$ and
$E_\omega=\sqrt{\omega^2+\Delta^2}$. The function $\delta f_3$ as
well as the functions $f_{0,1}$ are assumed to be small. In the F
layers all the components of the condensate function $\fcheck$ are
small. The functions $\ghat_0$ and $\ghat_3$ in Eq. (\ref{gfunc})
in the superconductor are given by
\begin{eqnarray}
\ghat_3&=&\sigmac(\tilde{g}_S+\delta g_0)+\sigmat g_3\\
\ghat_0&=&\sigmad g_2\; .
\end{eqnarray}
Here $\tilde{g}_S=\sgn\omega.g_S$, $g_S=|\omega|/E_\omega$. From the normalization condition
\begin{equation}
 \gcheck^2=1 \label{normalization}
\end{equation}
we obtain expressions relating the functions
$\delta g_0$, $g_{2,3}$ to the functions $\delta f_3$, $f_{0,1}$
\begin{equation}
\delta g_0=(f_S/\tilde{g}_S)\delta f_3,\;\; g_3=(f_S/\tilde{g}_S)f_0,
\;\; g_2=(f_S/\tilde{g}_S)f_1\; .
\label{relations}
\end{equation}
Now we linearize Eq. (\ref{Usad}) with respect to
$\delta\fcheck=i\taud(\sigmat\delta f_3+\sigmac f_0)+
i\tauu\sigmau f_1$ and obtain
\begin{equation}
\partial _{xx}^{2}\mbox{$\delta \fcheck$}-\kappa_S^2
\mbox{$\delta \fcheck$}=0  \label{UsadS}
\end{equation}
in the S layer, and
\begin{equation}
\partial _{xx}^{2}\mbox{$\delta \fcheck$}-\kappa_\omega^2
\mbox{$\delta \fcheck$}%
+i\kappa_J^2 \{\mbox{$\hat{\tau}_0$}[\mbox{$\hat{\sigma}_3$} ,%
\mbox{$\delta \fcheck$}]_{+}\cos \alpha \pm\mbox{$\hat{\tau}_3$}[%
\mbox{$\hat{\sigma}_2$},\mbox{$\delta \fcheck$}]_{-}\sin \alpha \}=0
\label{UsadF}
\end{equation}
in the F layers. Here $\kappa_S^2=2E_\omega/D_S$,
$\kappa_\omega^2=2|\omega|/D_F$, $\kappa_J^2=J\sgn\omega/D_F$
and $[A,B]_\pm=AB\pm BA$. The signs $\pm$ in Eq. (\ref{UsadF})
correspond to the right and left layer respectively.
The corresponding linearized boundary conditions for
$\delta \fcheck$ are
\begin{eqnarray}
(\gamma g_S)\partial_x\mbox{$\check{f}_F$}&=
&\partial_x\mbox{$\delta\check{f}_S$}
\label{bc1l} \\
\pm\gamma_b\xi_J\partial\mbox{$\check{f}_F$}&=
&-(\fcheck_S+\delta\fcheck_S)+g_S\fcheck_F\; ,  \label{bc2l}
\end{eqnarray}
where $\fcheck_S=i\taud\sigmat f_S$ and the signs $\pm$
correspond to the right and left layer.
Solutions for Eqs. (\ref{UsadS}-\ref{UsadF}) can be written
as a sum of exponential functions $\exp(\pm\kappa x)$,
where the $\kappa$'s are the eigenvalues of
Eqs. (\ref{UsadS}-\ref{UsadF}). In the S layer the
equations for $\delta f_3$,$f_{0,1}$ are decoupled
and there is only one eigenvalue $\kappa=\kappa_S$.
In the F-layers the equations are coupled and there
are three different eigenvalues \cite{VBE}
\begin{eqnarray}
\kappa _{1,2} &\equiv &\kappa _{\pm }\simeq\xi _{J}^{-1}
(1\pm i)\;,  \label{evpm} \\
\kappa _{3} &\equiv &\kappa _{\omega }=\sqrt{2|\omega |/D_{F}}\;. \label{evF}
\end{eqnarray}
We see from these equations that two completely
different lengths ${\xi _{J}}$ and ${\xi _{T}}$ determine the
decay of the condensate in the F layers. At all temperatures
$T<T_c^*$ the length ${\xi _{T}}$ much exceeds ${\xi _{J}}$ and is
the same the length describing the decay of the standard singlet
condensate in a normal metal.

We have assumed that $J\gg T_c^*$, which is realistic unless the
exchange field is extremely small. In order to find analytical
expressions for the functions $f_i$ we also assume that the
thicknesses of the S and F layers satisfy the conditions
\begin{equation}
d_S\ll\xi_S=\sqrt{D_S/2\pi T_c^*},\;\;\; d_F\gg \xi_J\;.\label{cond_ds}
\end{equation}
In this case the solutions for Eqs. (\ref{UsadS}-\ref{UsadF}) have the form
\begin{eqnarray}
\delta f_3(x)&=&a_3\cosh(\kappa_S x)\label{gen_S}\\
f_0(x)&=&a_0\cosh(\kappa_S x)\\
f_1(x)&=&a_1\sinh(\kappa_S x)\;,
\end{eqnarray}
in the S layer and
\begin{eqnarray}
f_{1}(x) &=&b_{1}\frac{\cosh\kappa_\omega(x-d_S-d_F)}{\cosh(\kappa_\omega d_F)}
+\mbox{${\rm sgn}$}\omega
\sin \alpha \left[ -b_{3+}e^{\kappa _{+}(x-d_{s})}+b_{3-}e^{-\kappa
_{-}(x-d_{s})}\right] \;,  \label{gen_f1} \\
f_{0}(x) &=&-\tan \alpha\; b_{1}\frac{\cosh\kappa_\omega(x-d_S-d_F)}
{\cosh(\kappa_\omega d_F)}+%
\mbox{${\rm sgn}$}\omega \cos \alpha \left[ -b_{3+}e^{-\kappa
_{+}(x-d_{s})}+b_{3-}e^{-\kappa _{-}(x-d_{s})}\right] \;,  \label{gen_f0} \\
f_{3}(x) &=&b_{3+}e^{-\kappa _{+}(x-d_{s})}+b_{3-}e^{-\kappa
_{-}(x-d_{s})}\;  \label{gen_f3}
\end{eqnarray}
in the right F layer. The solutions in the left F layer can be
easily obtained recalling that the function $f_1(x)$ is odd and
$f_{0,3}(x)$ are even functions  of $x$. From Eqs.
(\ref{gen_S}-\ref{gen_f3}) and the boundary conditions Eqs.
(\ref{bc1l}-\ref{bc2l}) we find
\begin{eqnarray}
\mbox{$\tilde{b}$}_{3\pm}=&b_{3\pm}(g_s+\gamma_b\xi_J\kappa_\pm)&=f_S\frac{\ktilde_S
\tanh\Theta_S M_{\mp}}{M_+ T_-+M_-T_+}  \label{btildepm} \\
\mbox{$\tilde{b}$}_1=&b_1(g_S+\gamma_b\xi_J\kappa_\omega\tanh\Theta_F)=&-f_S\sin\alpha\frac{%
\ktilde_S^2(\mbox{$\tilde{\kappa}$}_+-\mbox{$\tilde
{\kappa}$}_-)\sgn\omega}{%
\cosh^2\Theta_S\left( M_+ T_-+M_-T_+\right)}\; ,  \label{btilde1}
\end{eqnarray}
%%%%%%%%%%%%%%%%%%%%%%%%%%%%%%%%%%%%%%%%%%%%%%%%%%%5
 %%%%%%%%%%%%%%%%%%%%%%%%%%%%%%%%%%%%
 where
$\Theta_S=\kappa_sd_S$, $\Theta_F=\kappa_\omega d_F$,
$\ktilde_\pm=\kappa_\pm/(g_S+\gamma_b\xi_J\kappa_\pm)$,
$\ktilde=\kappa_\omega/(g_S+\gamma_b\xi_J\kappa_\omega\tanh\Theta_F)$,
$\ktilde_S=\kappa_S/(g_S\gamma)$ and
\begin{eqnarray*}
M_{\pm}&=&T_\pm(\ktilde_S\coth\Theta_S+\ktilde\tanh\Theta_F)+\tan^2\alpha\, C_\pm(\ktilde_S\tanh\Theta_S+\ktilde\tanh\Theta_F)\\
 T_\pm&=&\ktilde_S\tanh\Theta_S+ \ktilde_\pm\\
C_\pm&=&\ktilde_S\coth\Theta_S+\ktilde_\pm\;.
\end{eqnarray*}

\begin{figure}
\epsfysize= 5cm \vspace{0.2cm}
\centerline{\epsfbox{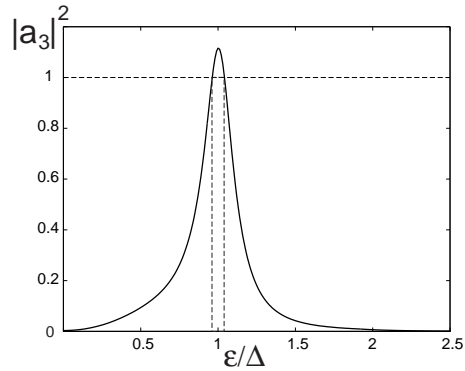}} \vspace{0.2cm}
\caption{Dependence of $|a_3^R|^2$ on the energy $\epsilon$. The dashed vertical lines show
the region in which our approach fails. Here $\gamma=0.05$,
$J/\Delta=25$, $d_S/\xi_\Delta=0.4$, $d_F/\xi_\Delta=0.5$,
$\gamma_b=0.5$, $\alpha=\pi/4$ and the damping factor $\Gamma=0.1$.
We have defined $\xi_\Delta=\sqrt{D_S/\Delta}$, where $\Delta$ is
the BCS order parameter.} \label{validity}
\end{figure}

%%%%%%%%%%%%%%%%
The solutions presented above are valid if the correction $\delta
f_3$ to the condensate function $f_s$ in the S layer is small (in
the F layer $\delta f_3$ is even smaller ). From Eqs.
(\ref{bc1l}-\ref{bc2l}) one can readily see that the condition
\begin{equation}
 \delta f_3(d_S)\sim\delta f_3(0)=a_3\cosh\Theta_S=\btilde_{3+}
 +\btilde_{3_-}-f_S\ll1\label{valid}
\end{equation}
should be satisfied. Here $|\Theta_S|\ll1$ is implied. Actually we have neglected the term $\delta f_3^2$ in the normalization condition (\ref{normalization}) assuming that $\delta f_3^2\ll1$ (see Fig. \ref{validity}). 

The amplitude $a_3$ of the SC depends on many parameters, such as
temperature (energy), $\gamma_b$, etc. Therefore, the validity of
our approach should be checked for every set of parameters. {\  If we are interested in  thermodynamical quantities such  as the critical temperature or the Josephson current, we may set $\omega\sim{\rm max}\{T,\Delta\}$. When calculating the density of states the situation is different because $f_S(\epsilon)$ has a singularity at $\epsilon=\Delta$ which is rounded off by a damping factor in the quasiparticle spectrum. In this case our approach breaks down near the energy $\epsilon\sim\Delta$ (see Fig.\ref{validity}), when the condition (\ref{valid}) is violated.}
 It is also clear that our approach is valid
provided either the temperature is close to the critical
temperature $T_c^*$ of the system  or $\gamma_b$ is not too small.
%%%%%%%%%%%%%%%
%%%%%%%%%%%%%%%%%

\begin{figure}
\epsfysize= 5cm
\vspace{0.2cm}
\centerline{\epsfbox{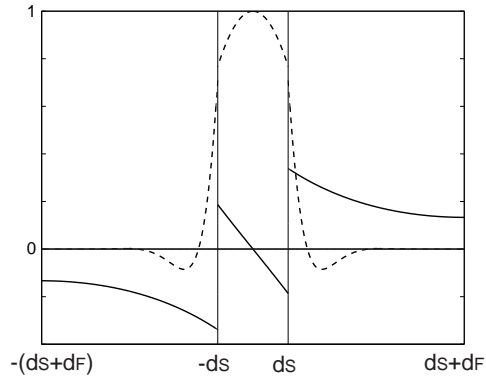}}
\vspace{0.2cm}
\caption{The spatial dependence of ${\rm Im}$(SC) (dashed line)
and the long-range part of ${\rm Re}$(TC) (solid line).
We have chosen $\gamma=0.2$, $J/T_C=50$, $\gamma_b=0.05$,
$d_F\sqrt{T_C/D_S}=2$, $d_S\sqrt{T_C/D_S}=0.4$ and $\alpha=\pi/4$.
The discontinuity of the TC at the S/F interface is because
the short-range part is not shown in this figure.}
\label{sp_dp}
\end{figure}

Now we discuss the properties of the obtained solutions (Eqs.
(\ref{gen_S}-\ref{btilde1})). From Eqs.
(\ref{gen_f3}-\ref{btildepm}) one can see that the SC is an even
function of $\omega$ and decays sharply in the ferromagnet over
the short distance $\xi_J$. In contrast, the amplitudes of the TC
$f_0$ and $f_1$ are odd functions of $\omega$ and penetrate the
ferromagnet over the longer distance $\xi_T=\sqrt{D_F/2\pi T}$.
The long-range part of TC determined by the amplitude $b_1$ has
the maximum at $\alpha=\pi/4$. This value of $\alpha$ corresponds
to a perpendicular orientation of the magnetizations in the F
layers. For a parallel ($\alpha=0$) or antiparallel alignment of
the magnetizations ($\alpha=\pi/2$) this amplitude decays to zero.
In Fig. \ref{sp_dp} we plot the spatial dependence of the SC and
the long-range part of the  TC. We see that both  amplitudes
are comparable at the S/F interface but the SC decays faster than
the TC.
%%%%%%%%%%%%%%%
%%%%%%%%%%%%%%%%%

The long-range part of TC leads to interesting observable effects
that will be discussed in the next sections.  In Refs.
\cite{BVE2,Swed} the conductance of a ferromagnetic wire attached
to a superconductor was calculated. It was assumed that the F wire
had a domain wall located at the S/F interface. This inhomogeneity
of the magnetization induces a TC, which leads to an increase of
the conductance for temperatures below $T_c$.

\subsection{Critical temperature}

In this section we discuss briefly the  effect of the TC on the
critical temperature $T_c^*$ of the structure. For the parallel
and antiparallel alignment of the magnetizations the critical
temperature of the multilayered structure $T_c^*$ was calculated
in many papers \cite{TheorTc,golubov}. The angle dependence of the
critical temperature in a F/S/F structure was analyzed in
Ref.\cite{buzdin}. However the form of the condensate function
presented in  Ref.\cite{buzdin} is not correct because the authors
started from an equation  different from Eq. (\ref{Usad}). As a
result, the long-range TC was completely lost.

The equation that determines $T_c^*$ has the form (we assume that
$d_S\ll\xi_S$, see Refs.\cite{TheorTc})
\begin{equation}
\log\left(\frac{T_c}{T_c^*}\right)=2\pi T_c^*\sum_{\omega=-\infty}^\infty
\left\{ \frac{1}{\omega}-i\frac{\btilde_{3p}+\btilde_{3-}}{\Delta}\right\}\; .
\label{tc}
\end{equation}
We have obtained a solution for $\btilde_{3\pm}$, (Eq.
(\ref{btildepm})), assuming that $\Delta$ is constant in space (
this approximation corresponds to the so-called single-mode
approximation used in many earlier works \cite{TheorTc}). It is
established in Ref.\cite{golubov} that for some parameters this
approximation gives a rough estimate for $T_c^*$. A careful
analysis of Ref.\cite{golubov} shows that  $T_c^*$ remains finite
even for values of the parameters $\gamma$, $\gamma_b$,
$\kappa_J$, for which other approaches predict a zero critical
temperature.  We will not discuss quantitatively the dependence of
$T_c^*$ on the angle $\alpha$. Note however that, as follows from
Eqs. (\ref{btildepm}) and (\ref{tc}), the critical temperature
$T_c^*$ depends on $\alpha$ and $d_F$ even in the case when
$d_F\gg\xi_J$ (if $\alpha\neq0$). This dependence is due to the
long-range part of the TC and, in order to determine it, one has,
generally speaking, to go beyond the single-mode approximation. Note, however, that this dependence may be weak.

\subsection{Local density of states}

In this section we calculate the change of the local DoS in the F layers
due to the TC. It is clear that, for distances from the S/F
interface larger than $\xi_J$, only the TC leads to a variation of
the local  DoS. Thus, if the thickness $d_F$ is much larger than $\xi_J$
one can detect directly the presence of the TC performing
measurements of the DoS at the outer side of one of the F layers.
Any deviation from the normal value would be only due to the TC.

\begin{figure}
\epsfysize= 6cm
\vspace{0.2cm}
\centerline{\epsfbox{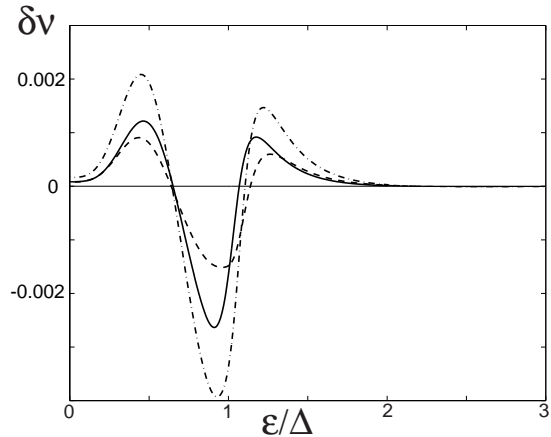}}
\vspace{0.2cm}
\caption{The normalized DoS $\delta\nu$ as a function
of the energy for $\alpha=3\pi/8$ (solid line),
$\alpha=\pi/8$ (dashed line) and $\alpha=\pi/4$
(point-dashed line). Note that for $\alpha=0,
\pi/2$ $\delta\nu=0$. We have chosen $\gamma=0.05$,
$J/\Delta=25$, $\gamma_b=0.5$, $d_F/\xi_\Delta=0.5$,
and $d_S/\xi_\Delta=0.4$. Here $\xi_\Delta=
\sqrt{D_S/\Delta}$ and $\Delta$ is the BCS order parameter.}
\label{dos_a}
\end{figure}

\begin{figure}
\epsfysize= 6.5cm
\vspace{0.2cm}
\centerline{\epsfbox{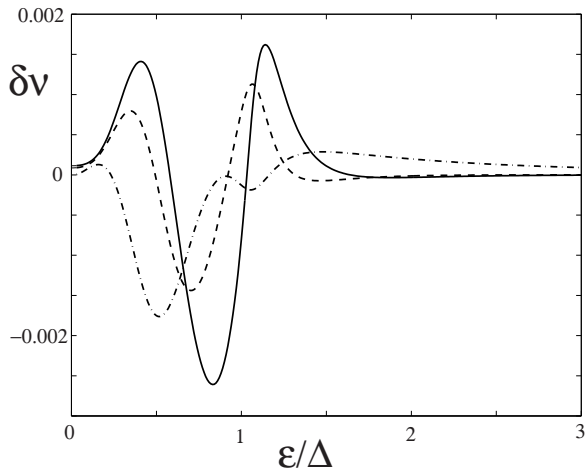}}
\vspace{0.2cm}
\caption{The normalized DoS $\delta\nu$ as a function
of the energy for $d_F/\xi_\Delta=0.8$ (solid line),
$d_F/\xi_\Delta=1.2$  (dashed line).{\  The point-dashed line shows the contribution to the DoS from the SC ($f_3$). The latter is multiplied by a factor of 100.} We have chosen $\alpha=\pi/4$. All other parameters are the same as in Fig. \ref{dos_a}}
\label{dos_df}
\end{figure}

\begin{figure}
\epsfysize= 5cm
\vspace{0.2cm}
\centerline{\epsfbox{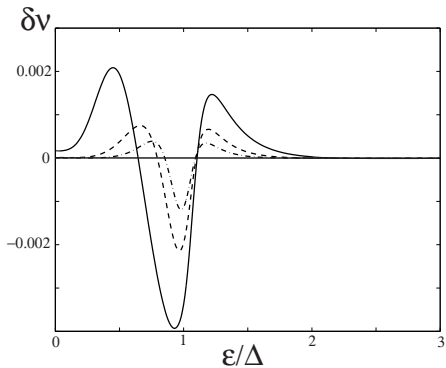}}
\vspace{0.2cm}
\caption{The normalized DoS $\delta\nu$ as a function of
the energy for $\gamma_b=0.5$ (solid line),$\gamma_b=1$
(dashed line) and $\gamma_b=1.5$ (point-dashed line).
We have chosen $d_F/\xi_\Delta=0.5$.  All other parameters are the same as in Fig. \ref{dos_df}}
\label{dos_gb}
\end{figure}

We calculate the local  DoS at $x=d_S+d_F$. The expression for the
normalized DoS is (we ignore the difference in the DoS for the up
and down spin directions. {\ This approximation is consistent with the quasiclassical assumption that $J\ll \epsilon_F$, where $\epsilon_F$ is the Fermi energy})
\begin{equation}
\tilde{\nu}=\frac{\nu}{\nu_0}=\frac {1}{8}{\rm Tr}\left(\taut\sigmac\right)
\left(\gcheck^R-\gcheck^A\right)\;, \label{dos_gen}
\end{equation}
{\ where $\nu_0$ is the DoS in the normal state, thus $\tilde{\nu}=1+\delta\nu$ ($\delta\nu$ is a correction due to the proximity effect).}
As it was mentioned before, in the case $d_F\gg\xi_J$ only
the TC (i.e. the functions $f_0(x)$ and $f_1(x)$) contributes
to the DoS. From the normalization condition Eq.
(\ref{normalization}) and Eq. (\ref{dos_gen}) we obtain
\begin{equation}
 \delta\nu=\frac{1}{2}{\rm Re}\frac{\left(b_1^R\right)^2}
 {\cos^2\alpha\cosh^2\Theta_F^R}\; ,\label{dos}
\end{equation}
where $\Theta_F^R=\sqrt{-2i\epsilon/D_F}d_F$, and $b_1^R$ is the
amplitude of the retarded Green's function in Eqs.
(\ref{gen_f1}-\ref{gen_f0}). It is obtained from $b_1$ by
replacing $\omega$ by $-i\epsilon$. In Figs. \ref{dos_a},
\ref{dos_df} and \ref{dos_gb} we plot the dependence of
$\delta\nu$ on $\epsilon$ for different $\alpha$, $d_F$ and
$\gamma_b$, respectively. For the range of parameters chosen in
these plots the function $|a_3(\epsilon)|^2$ has the shape shown in
Fig.\ref{validity}. Thus, our approach is valid almost for all
energies and fails only in a very narrow region close to
$\epsilon=\Delta$. In order to avoid singularities in $f_S^R$ we
have taken into account a finite damping factor $\Gamma=0.1$ in
the expression for $f_S^R$:
\begin{equation}
f_S^R=\frac{\Delta}{\sqrt{(\epsilon+i\Gamma)^2-\Delta^2}}\;.
\end{equation}
 As follows from Eq. (\ref{btilde1}) $\delta\nu$ is zero for
 $\alpha=0,\pi/2$. The largest change in the DoS is achieved
 when  $\alpha=\pi/4$ (perpendicular orientation of magnetizations
 in the F layers). We see that the correction to the DoS is small
 but observable. Kontos et al. presented in Ref.\cite{kontos}
 measurements of $\delta\nu$ in thin F layers (few nanometers).
 The order of magnitude of the observed $\delta\nu$
 ($\sim 10^{-3}$) is the same as the presented in
 Figs.\ref{dos_a}-\ref{dos_gb}. However,
 in Ref.\cite{kontos} the variation of the DoS was caused
 by the penetration of the SC into the F layer over the
 short distance $\xi_J$. In our case such a variation
 can be observed in much thicker F layers
 ($d_F\sim\xi_T=\sqrt{D_F/2\pi T_c^*}\gg\xi_J$).

 It is interesting to compare our result for the
 FSF structure with non-collinear magnetization with corresponding
 results for NSN structures (N is a normal layer). At first
 glance, the behavior of the odd triplet condensate in the
 ferromagnet is very similar to that of the conventional singlet
 condensate in a normal metal. In both cases the amplitude of the condensate
 decays exponentially with the length ${\xi_T}$ ( Eq. (\ref{evF})).
 However, there is an essential difference. In the N layer an
 energy gap is induced due to the singlet condensate. The value of
 the energy gap is determined by
 ${\rm min}\{\Delta,D_N/(\sigma_Nd_NR_b)\}$\cite{mcmill}.
 In contrast, no subgap appears in the ferromagnet due to the triplet
 odd condensate considered here, although the TC  penetrates over
 the entire F layer provided its thickness
 $d_F$ is not very large, $d_F\leq\xi_T$. The main reason for the absence of a subgap $\epsilon_{sg}$ in the FSF system is the following. In SN structures the condensate function is not small at energies $|\epsilon|\lesssim\epsilon_{sg}$. The exchange field shifts this energy interval by the large value $J$ so that  at low energies the condensate function ( both singlet and triplet) is small if $\gamma_b$ is not too small. {\ Note also that the amplitude of the TC is smaller than the amplitude of the SC in a NSN structure since it contains a large parameter $\kappa_\pm\sim \sqrt{J}$ in the denominator (see Eqs. (\ref{btildepm}-\ref{btilde1})).}

{\
For completeness we finally note that the change of the local DoS in the ballistic case ($J\tau\gg 1$)  was considered in Ref.\cite{BVEdos} and in the pure ballistic case ($\tau\rightarrow\infty$) in Ref.\cite{valls}. It turns out that the results in these two cases differ greatly from those obtained  in the present paper for a diffusive system ($J\tau\ll 1$).}
%%%%%%%%%%%%%%%
%%%%%%%%%%%%%%%%%%%%%%%%%%%%%%%%
%%%%%%%%%%%%%%%%%
%%%%%%%%%%%%%%%%%%%%%%%%%%%%%%%%
%%%%%%%%%%%%%%%%%

\section{Josephson Current in a F/S/F/S/F structure}
In this section we calculate the Josephson current between the S
layers of a FSFSF structure. We assume again that the thickness of
the F layers $d_F$ is much larger than $\xi_J$ (Eq.
(\ref{cond_ds})). In this case the Josephson coupling between the S
layers is due to the long range part of the TC. Therefore the
supercurrent in the transverse direction is unusual, since it is
caused by the triplet component of the condensate that is odd in
frequency and even in momentum.

At the same time, the in-plane superconductivity is caused mainly
by the ordinary singlet component. Therefore the macroscopic
superconductivity due to the Josephson coupling between the layers
is an interesting combination of the singlet superconductivity
within the layers and the odd triplet superconductivity in the
transversal direction.

\begin{figure}
\epsfysize= 5cm \vspace{0.2cm}
\centerline{\epsfbox{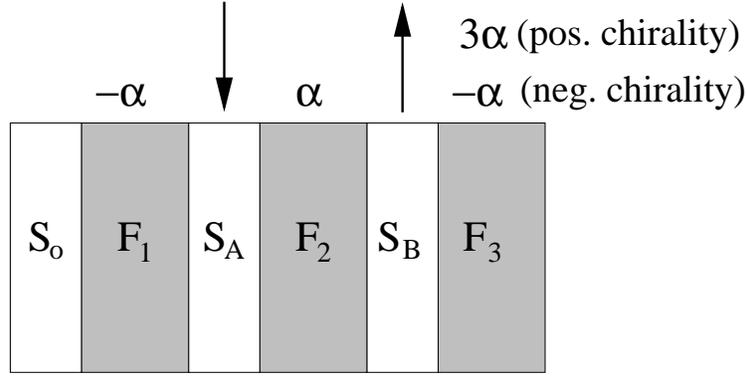}} \vspace{0.2cm}
\caption{The multilayered structure considered. The arrows show
the bias current. In the case of positive (negative) chirality the
magnetization vector ${\bf M}$ of the layer ${\rm F}_3$ makes an angle $3\alpha$
($-\alpha$) with the $z$- axis,{\  {\it i. e.} in the case of positive chirality the vector ${\bf M}$ rotates in one direction if we go over from one F layer to another whereas it oscillates in space in the case of negative chirality.}} \label{geom2}
\end{figure}

\begin{figure}
\epsfysize= 5cm
\vspace{0.2cm}
\centerline{\epsfbox{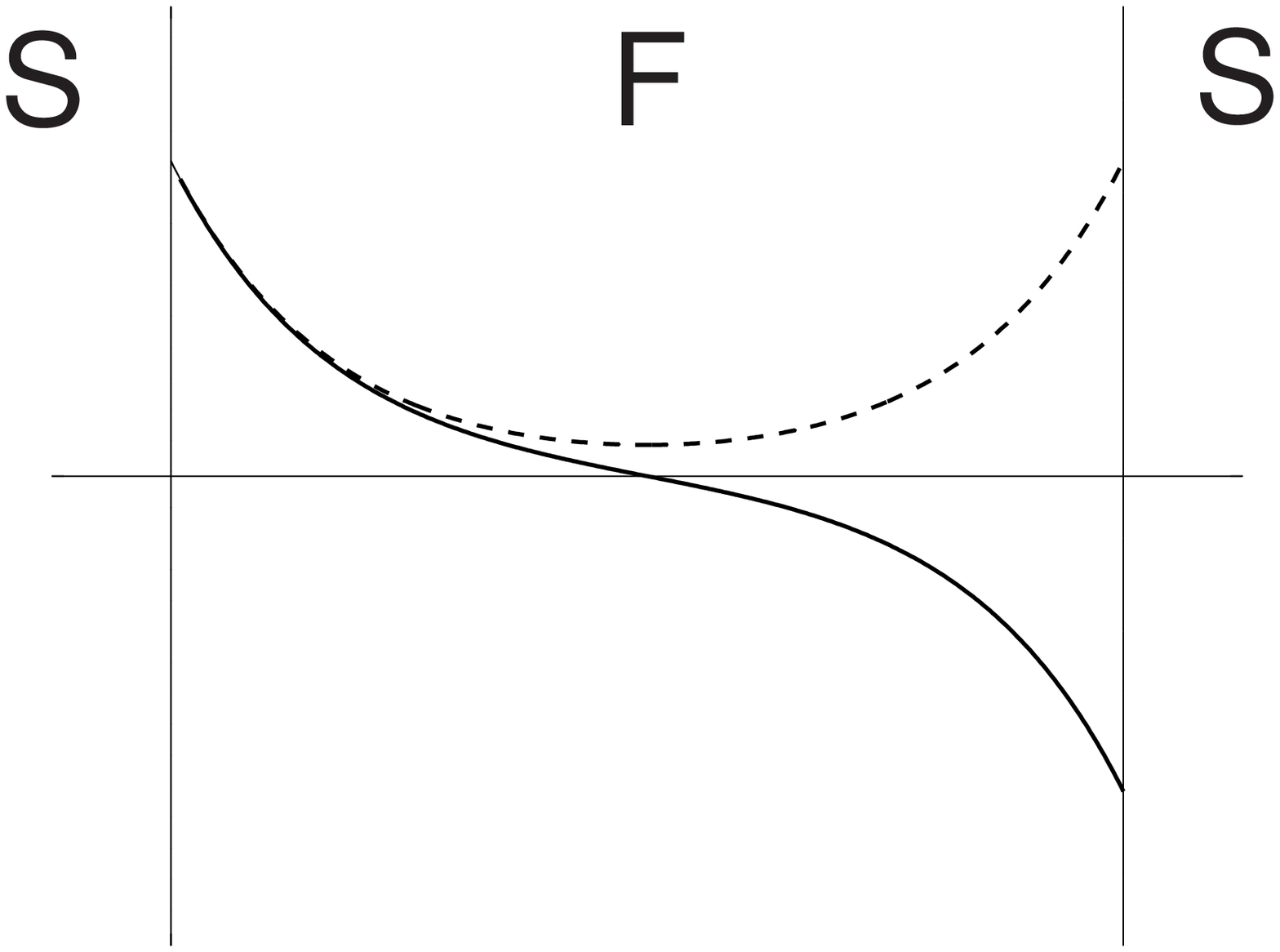}}
\vspace{0.2cm}
\caption{The spatial dependence of the amplitude of the TC $f_1(x)$ in the case of positive (solid line)and negative (dashed line) chirality. }
\label{sp_dp2}
\end{figure}

We will see that the unusual character of the superconductivity in
the transversal direction leads to peculiarities of the Josephson
effect. For example, if the bias current flows through the
terminal superconducting layer S$_{\rm O}$ and S$_{\rm A}$ (see
Fig.\ref{geom2}), the supercurrent is zero because of the different symmetry of the condensate in  S$_{\rm O}$ and S$_{\rm A}$. In order to observe
the Josephson effect in this structure the bias current has to
pass through the layers S$_{\rm A}$ and  S$_{\rm B}$, as shown in
Fig.\ref{geom2}. The supercurrent between $S_{\rm A}$ and $S_{\rm B}$ is
non-zero because each superconductor has its ``own'' TC and the
phase difference $\varphi$ is finite.
%%%%%%%%%%%%%%%%%%%%%%%%%%%%%%%%
%%%%%%%%%%%%%%%%%
The Josephson current $I_S$ is given by the expression
\begin{equation}
I_S=(L_yL_z)\sigma_F{\rm Tr}\left(\taut\sigmac\right)
\sum_\omega\check{f}\partial_x\fcheck\; \label{gen_current}
\end{equation}
This current was calculated for the case of small angles $\alpha$
in Ref. \cite{VBE}. Here $L_yL_z$ is the area of the interface and
$\sigma_F$ is the conductivity of the F layer. The simplest way to
calculate the $I_S$ is to assume a weak coupling between the S
layers, which corresponds to the case when the condition
$d_F>\xi_T$ holds. In this case the long-range part of the TC is
given by the sum of two terms each of those is induced by the
layers  S$_{\rm A}$  and S$_{\rm B}$ in Fig. \ref{geom2}:
\begin{equation}
\fcheck(x)=\fcheck_{\rm A}(x)+\check{S}.\check{U}.
\fcheck_{\rm B}(x-d_S-d_F)\check{U}^+.\check{S}^+\; , \label{weak_coup}
\end{equation}
where
\begin{equation}
\fcheck_{\rm A}(x)=e^{-\kappa_\omega(x-d_S)}
\left(b_1.i\tauu\sigmau+b_0.i\taud\sigmac\right)
\end{equation}
is the long range part of the TC induced by the layer S$_{\rm A}$.
The coefficient $b_1$ is given by  Eq.(\ref{btilde1}) and
$b_0=-\tan\alpha\, b_1$. If the S$_{\rm A,B}$/F interfaces are
identical as well as the superconductors S$_{\rm A}$ and  S$_{\rm
B}$ , the function $\fcheck_{\rm B}$ is equal to $\fcheck_{\rm A}$
if one replace the exponential function $\exp(-\kappa_\omega(x-d_S))$
by $\exp(\kappa_\omega(x-d_S-d_F))$. The phase of the S$_{\rm A}$ layer
is set to be zero and the phase of the  S$_{\rm B}$ is $\varphi$.
This phase has been taken into account by the gauge transformation
performed with the help of the matrix
$\check{S}=\tauc\cos(\varphi/2)+i\taut\sin(\varphi/2)$. The
magnetizations ${\bf M}$ of the layers F$_1$ and  F$_2$ make an
angle $\mp \alpha$ with the $z$-axis respectively. For the
direction of ${\bf M}$ in the   F$_3$ we consider two cases: a)
the direction of magnetization is $-\alpha$ (negative chirality)
or  b) $2\alpha$ (positive chirality). In the  latter case the
matrix $\check{U}$ in Eq. (\ref{weak_coup}) is given by
\begin{equation}
\check{U}=\tauc\sigmat\cos\alpha+i\taut\sigmad\sin\alpha\;.
\end{equation}
In the case of negative chirality, $\check{U}$ is the unit matrix
and one has to change the sign of $\alpha$ in the expression for
the function $\fcheck_{\rm B}$ (Eq. (\ref{weak_coup})). In Fig. \ref{sp_dp2} we show schematically the spatial dependence of $f_1(x)$.
%%%%%%%%%%%%%%%%%%%%%%%%%%%%%%%%
%%%%%%%%%%%%%%%%%

Substituting  Eq. (\ref{weak_coup}) into
Eq. (\ref{gen_current}) one obtains after simple
transformations $I_S=I_c\sin\varphi$, where
\begin{equation}
eR_FI_c=\pm 2\pi T\sum_\omega\kappa_\omega d_Fb_1^2(\alpha)
\left(1+\tan^2\alpha\right)e^{-d_F\kappa_\omega}\; ,\label{current}
\end{equation}
where $b_1(\alpha)$ is given in Eq. (\ref{btilde1}) and
the sign ``+'' (``-'') corresponds to the positive (negative)
chirality. In the case of negative chirality the critical current
is negative ($\pi$-contact). It is important to emphasize that the
nature of the $\pi$-contact differs from that predicted in
Refs.\cite{PiContact} and observed in Ref.\cite{Ryazan}. In our
case the negative Josephson coupling is due to the TC and can be
realized in S/F structures with negative chirality. This gives a
unique opportunity to switch experimentally between the $0$ and
${\pi}$-contacts by changing the angles of the mutual
magnetization of the layers. It is worth mentioning that another effect concerning the  chirality of the ${\bf M}$ vector was studied by the authors in Ref.\cite{DW}. It was shown that the resistance of a multi-domain ferromagnetic wire depends on the chirality of the ${\bf M}$ variation in space.

\begin{figure}
\epsfysize= 5cm
\vspace{0.2cm}
\centerline{\epsfbox{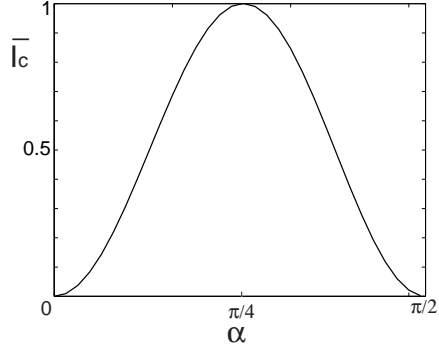}}
\vspace{0.2cm}
\caption{Dependence of the critical current
( normalized with respect to the maximum value)
on the angle $\alpha$. We have chosen the same
values as in Fig. \ref{dos_a}}
\label{current_a}
\end{figure}

In Fig. \ref{current_a} we plot the dependence of $I_c$ on the
angle $\alpha$. If the orientation of ${\bf M}$ is parallel ($\alpha=0$) or antiparallel ($\alpha=\pi/2$) the amplitude of the triplet component is zero and therefore there is no coupling between the neighboring S layers, i.e. $I_c=0$.
For any other angle between the magnetizations the amplitude of the TC is finite. This leads to a non-zero critical current. At $\alpha=\pi/4$ ( perpendicular orientation of ${\bf M}$) $I_c$ reaches its maximum value.
%%%%%%%%%%%%%%%%%%%%%%%%%%%%%%%%
%%%%%%%%%%%%%%%%%
The weak coupling assumption ($d_F>\xi_T$) leads to
an exponential decay of $I_c$ with increasing $d_F$ ( Eq. (\ref{current})).
In the case $d_F\leq\xi_T$, Eq. (\ref{current}) is not valid. One
can easily obtain $I_c$ for the case of an arbitrary $d_F$ and
small $\alpha$. It turns out that in this case Eq. (\ref{current})
remains valid if the exponential factor $\exp(-\kappa_\omega d_F)$ is
replaced by $\cosh^{-2}(\kappa_\omega d_F/2)$ and in the expression for
$b_1$ (Eq. (\ref{btilde1}))  $\Theta_F$ is replaced  by
$\Theta_F/2$.

In order to estimate the value of the  critical current $I_c$,
we use Eq. (\ref{current}). If $d_F$ exceeds
the length $\xi_T$ ( for example $d_F/\xi_T=2$) only
the term with $n=0$ (i.e. $\omega=\pi T_c^*$) is important
in the sum. In this case one obtains
\begin{equation}
\frac{eR_FI_c}{T_c^*}=\frac{4}{\pi}\left(\frac{\Delta}
{T_c^*}\right)^2e^{-\kappa_Td_F}C\; , \label{current2}
\end{equation}
where the factor $C$ can be easily expressed in terms of $M_\pm$,
$T_\pm$, etc. Thus, $C$ depends on many parameters such as
$\gamma$, $\gamma_b$, $\kappa_J$, etc. We estimate $C$ for values
of these parameters similar to those which were used in Ref.
\cite{golubov}: $\gamma_b=0.5$, $\gamma=0.1$, $d_S\kappa_S=0.4$,
$d_F\kappa_\omega=1.5$, $\kappa_\omega/\kappa_S=3$. We get
$C=10^{-2}-10^{-3}$ for $\kappa_Jd_S=5-10$. The expression
(\ref{current2}) for $I_c$  also contains the  parameters
$(\Delta/T_c^*)^2$ and $\exp(-d_F\kappa_T)$ which are also small.
We  note however that if $d_F\leq\xi_T$, the exponential function
is replaced by a numerical factor of the order of 1. The factor
$(\Delta/T_c^*)^2$ is also of the order 1 if the temperature is
not close to $T_c^*$. Taking $\sigma_F^{-1}=$60$\mu\Omega$.cm
({\it cf.} Ref.\cite{golubov}) and $d_F\sim\xi_T\sim$200 nm we
obtain $I_c\sim 10^4-10^5$ A.cm$^{-2}$; that is, the critical
current is a measurable quantity ( see experimental
works\cite{Ryazan}) and the detection of the TC is possible.

%%%%%%%%%%

%%%%%%%%%

\section{Effect of Spin-Orbit interaction}

So far the only interaction we have considered in the ferromagnet
is the exchange field $J$ acting on the conducting electrons.
However, in reality spin-orbit interactions that appear due to
interactions of electron spins with spin orbital impurities may
become important. {\ Following again the notation of Ref.\cite{BVE3} we write an  additional
term in the Hamiltonian which describes the spin-orbit part as\cite{alexander,demler}}
\begin{equation}
H_{so}=\frac{U_{so}}{2p_F^2}\sum_{n,s,p,n',s',p'}c^+_{nsp}\left(\bold{p}\times\bold{%
p^{\prime}}\right)(\check{S})_{ss'}^{nn'}c_{n's'p'}\; ,
\end{equation}
where $\check{S}=(\mbox{$\hat{\sigma}_1$},\mbox{$\hat{\sigma}_2$},%
\mbox{$\hat{\tau}_3$}\mbox{$\hat{\sigma}_3$})$ and $\bold{p}$ and $\bold{%
p^{\prime}}$ are the momenta before and after scattering at the
impurities. Although in general the characteristic energy of the
spin-orbit interaction is much smaller than the exchange energy,
it can be comparable with the superconducting gap $\Delta$ and
therefore this effect should be taken into account when describing
the supercurrent.

In the Born approximation the self-energy is given by
\begin{equation}
\check{\Sigma}_{so}=n|U_{so}|^2<G>_{s.o.}\;,\;\; {\rm where} \;\;
<G>_{s.o.}=\nu\int d\xi_p\int \frac{d\Omega}{4\pi}\left(\bold{n}\times\bold{%
n^{\prime}}\right)\check{S}G\check{S}\left(\bold{n}\times\bold{n^{\prime}}%
\right)\; .
\end{equation}
Here $\bold{n}$ is a unit vector parallel to the momentum.
Including this term in the quasiclassical equations is
straightforward and the resulting Usadel equation takes the form
\cite{alexander}

\begin{equation}
-iD\partial_{{\bf r}}(\mbox{$\check{g}$}\partial_{{\bf r}}\mbox{$\check{g}$}%
)+i\left(\mbox{$\hat{\tau}_3$}\partial_t\mbox{$\check{g}$}%
+\partial_{t^{\prime}}\mbox{$\check{g}$}\mbox{$\hat{\tau}_3$}\right)+ \left[%
\check{\Delta},\mbox{$\check{g}$}\right]+J\left[\check{n},\mbox{$\check{g}$}%
\right]+\frac{i}{\tau_{s.o.}}\left[\check{S}\mbox{$\hat{\tau}_3$}%
\mbox{$\check{g}$}\mbox{$\hat{\tau}_3$}\check{S},\mbox{$\check{g}$}\right]%
=0\; ,  \label{spin_orbit_usadel}
\end{equation}
where
\begin{equation}
\frac{1}{\tau_{s.o.}}=\frac{1}{3}\nu n\pi\int\frac{d\Omega}{4\pi}
|U_{so}|^2\sin^2\theta\;\label{tos}
\end{equation}
is the spin-orbit scattering time.

As before, one can linearize Eq. (\ref{spin_orbit_usadel}) in the
F-layer and obtain equations for the condensate function
$\mbox{$\check{f}$}$ similar to Eqs. (\ref{UsadS}-\ref{UsadF}) but
now including the spin-orbit interaction term. The solution again has
 the form
\begin{equation}
\mbox{$\check{f}$}(x)=i\mbox{$\hat{\tau}_2$}\otimes(f_0(x)%
\mbox{$\hat{\sigma}_0$}+f_3(x)\mbox{$\hat{\sigma}_3$})+i\mbox{$\hat{\tau}_1$}%
\otimes f_1(x)\mbox{$\hat{\sigma}_1$}\; .
\end{equation}
The functions $f_i(x)$ are given by $f_i(x)=\sum_jb_j\exp[\kappa_jx]$, where
the new eigenvalues $\kappa_j$ are
\begin{eqnarray}
\kappa^2_{\pm} &=&\pm \frac{2i}{D_F}\sqrt{J^2-\left(\frac{4}{\tau_{so}}%
\right)^2}+\frac{4}{\tau_{so}D_F} \\
\kappa^2_0&=&\kappa_\omega^2+2\left(\frac{4}{\tau_{s.o.}D_F}\right)\; .
\end{eqnarray}
We see from these equations that the singlet and triplet
components are affected by the spin-orbit interaction making the
decay of the condensate in the ferromagnet faster. In the limiting
case $4/\tau_{so}>J,T_c$ both the components penetrate
over the same distance $\xi_{s.o.}=\sqrt{\tau_{so}D_F%
}$ and therefore the long-range effect is suppressed. In this case
the characteristic oscillations of the singlet component are
destroyed \cite {demler}. In the more interesting case
$4/\tau_{so}\sim T_c<J$, the singlet component is not affected and
penetrates over distances of the order $\xi_J$. At the same time,
the triplet component is more sensitive to the spin-orbit
interaction and the penetration length equals
min($\xi_{so},\xi_T$)$>\xi_J$.

{\ Spin-orbit interaction is relevant  in systems with large Z elements. The characteristic spin-orbit energy $1/\tau_{s.o.}$ also depends on scattering concentration and density of states ({\it cf.} Eq. (\ref{tos})). Experimental dataconcerning this energy  is still unclear and controversial, mainly due to the difficulty to separate the contribution of the spin-orbit from other scattering types. From numerical band structure calculations one can estimate the parameter $J\tau_{s.o}$. For example, for a typical magnetic transition metal, like Fe, in the dirty limit $J\tau_{s.o}\sim 10^2$, while for dirty Gd   $J\tau_{s.o}\sim 10$ ( see Ref.\cite{demler2} and references therein). Thus, according to our model, material like transition metals are better candidates in order to observe the predicted effects.}

Thus, provided the spin-orbit interaction is not very strong, the
penetration of triplet condensate over the long distances
discussed in the preceding sections is still possible, although
the penetration length is  reduced.

\section{Conclusion}
We studied odd, s-wave, triplet superconductivity that may arise
in S/F multilayered structures with a non-collinear orientation of
magnetizations.

It was assumed that the orientation of the magnetization is not
affected by the superconductivity (e.g. the energy of the magnetic
anisotropy is much larger than the superconducting energy). The
analysis was carried out in the dirty limit ($J\tau\ll 1$) when
the  Usadel equation is applicable.

It was shown that for all values of $\alpha$ the condensate
function consists of a singlet (SC) and a triplet (TC) components.
Even in the case of a homogenous magnetization ($\alpha=0$), in
addition to the SC, the TC with the zero projection onto the $z$
axis arises. In this case, both the SC and the TC decay in the F
layers over a short distance given by $\xi_J=\sqrt{D_F/J}$. If the
magnetization vectors ${\bf M}$ are not collinear $\alpha\neq
0,\pi/2$, all projections of the TC appear, in particular, those
with non-zero projection on the $z$-axis. In this case, the TC
penetrates  the F layer over a long distance
$\xi_T=\sqrt{D_F/2\pi T}$.  In the presence of spin-orbit
interaction this penetration length is given by
min($\xi_{so},\xi_T$), where $\xi_{so}=\sqrt{\tau_{so}D_F}$.
Generally, this length  may be much larger than $\xi_J$.

Thus, if the condition $d_F\gg\xi_J$ is fulfilled the Josephson
coupling between neighboring S layers is only due to the TC.
Therefore in this case a new type of superconductivity may arise
in the multilayered structures with non-collinear magnetizations.
The supercurrent within each S layer is caused by the SC, whereas
the supercurrent across the layers is caused by the triplet
condensate, which is odd in the frequency $\omega$ and even in the
momentum.

The TC in our case is completely different from the triplet
condensate found in Sr$_2$RuO$_4$ \cite{SrRuO}. In the latter case
one has  a $p$-wave, even in $\omega$, triplet superconductivity,
which is suppressed by impurity scattering. In contrast, the TC we
have considered is not affected by non-magnetic impurities. The
reason for the existence of the long-range TC is the fact that if
$\alpha\neq0$, the SC and the TC are coupled and, in addition to
$\kappa_\pm=\xi_J^{-1}(1\pm i)$, the eigenvalue
$\kappa_T=\xi_T^{-1}$ appears. The latter corresponds to the
long-range penetration of the TC in the ferromagnet.

The triplet superconductivity in S/F structures possesses  an
interesting property: the Josephson current depends on the
chirality of the magnetization ${\bf M}$: If the ${\bf M}$ vector
rotates in only one direction (the positive chirality) the
critical current $I_c$ is positive. If the direction of the ${\bf
M}$ vector oscillates in space (the negative chirality) then
$I_c<0$. In the latter case spontaneously circulating currents
must arise in the structure. This result can be explained as
follows: if the chirality is positive the averaged ${\bf M}$
vector $<{\bf M}>$ is zero and the S/F structure behaves as a
superconductor with anisotropic properties (the singlet
superconductivity along the layers and the triplet
superconductivity across them). In the case of the negative
chirality the average in space yields a non-zero magnetization
$<{\bf M}>\neq0$. In such a superconductor with a build-in
magnetic moment the circulating currents arise as they arise in
superconductors of the second type in the mixed state.

Note also that in a single Josephson FSFSF junction a non-zero
magnetic field exists also inside the junction and this causes
Meissner currents. However, the experiment of Ref. \cite{Ryazan}
on SFS junctions shows that the observed Fraunhofer pattern
corresponds to $<{\bf M}>=0$ in the F layer. This behavior
according to the authors of Ref.  \cite{Ryazan} may be attributed
to a multi-domain structure.

It would be interesting to carry out experiments on S/F structures
with non-collinear magnetization in order to observe this new type
of superconductivity. As follows from a semiquantitative analysis,
the best conditions to observe the Josephson critical current
caused by the TC are high interface transparency (small
$\gamma_b$) and low temperatures. These conditions are a bit
beyond our quantitative study. Nevertheless, all qualitative
features predicted here (angle dependence, etc) should remain in a
general case when one has to deal with the non-linear Usadel
equation.

Another type of experiments that may detect the triplet condensate
is measuring the local density of states. As we have shown in the
second section, the long-range TC may be detected by measuring the
local  DoS of the F layers.

We would like to thank SFB 491 and the German-Israeli Foundation (GIF)
for a financial support.

\bigskip
{\
{\  Note added:} After the submission of this manuscript a paper\cite{0303} appeared in which a detailed study of the critical temperature in FSF structure with noncollinear magnetizations in the F layers has been presented. 
}


\begin{references}
\bibitem{foot} The particle-hole space is often called the Nambu space due to the scheme introduced by Nambu. A closely related scheme was developed by Gor'kov prior to Nambu's work (see Ch. 7.2 in Ref.\cite{schr}).
\end{references}
\end{document}